%% file: preheat.tex
\documentstyle[psfig]{mn2e}
\input{macro1}

\begin{document}


\title[Accretion and Cooling of Preheated Gas]
      {The Accretion and Cooling of Preheated Gas in Dark Matter Halos}

\author[]
       {Yu Lu$^{1}$\thanks{E-mail: luyu@astro.umass.edu},
	H.J. Mo$^{1}$
\\
        $^1$ Department of Astronomy, University of Massachusetts,
        Amherst MA 01003-9305, USA}

\date{}
\maketitle
\label{firstpage}


\begin{abstract}
We use a one-dimensional hydrodynamical code to investigate 
the effects of preheating on gas accretion and cooling in 
cold dark matter halos. In the absence of radiative cooling, 
preheating reduces the amount of gas that can be accreted into a halo, and 
the accreted gas fraction is determined by the ratio of the initial 
specific entropy of the gas to the virial entropy of the halo, $\sph/\sv$. 
In the presence of radiative cooling, preheating affects the gas fraction that 
can cool in two different ways. For small halos with masses $M<10^{12}\hMsun$, 
preheating suppresses gas accretion, but most of the accreted gas can cool. 
For more massive halos, preheating affects the cold gas fraction 
not only by reducing the amount of accreted gas, but also by 
reducing the cooling efficiency. For both small and massive halos,  
gas cooling is delayed by preheating if the halo gas is assumed to be 
a single-phase medium. However, if the halo gas is assumed to be a multi-phase 
medium, cooling can occur over a wider range of redshifts. Unlike 
in a single-phase medium where cooling is inside-out, 
gas cooling in a preheated multi-phase medium can occur simultaneously  
over a wide range of radii. As examples, two specific preheating cases are 
investigated. In the first case, the preheating specific entropy is 
assumed to be proportional to the virial entropy of the halo, 
$\sph\propto \sv$, as expected from AGN feedback.  
Such preheating effectively suppresses radiative cooling in halos with 
$M>10^{13}\hMsun$, but has little effect on smaller halos. We suggest that this 
may be the reason why the stellar mass function of galaxies breaks sharply 
at the massive end. Such preheating also helps create the hot 
diffused halos within which the ``radio mode'' feedback of AGNs can act effectively. 
In the second case, the intergalactic medium is assumed to be warm. Here
the total amount of gas that can cool in a halo scales with halo mass 
as $\propto M^2$, as would be required to match the observed stellar- and 
HI-mass functions in the current CDM model at the small mass end.
\end{abstract}




\section{Introduction}
\label{sec:intro}
According to the current theory of structure formation, 
galaxies are believed to form in cold dark matter (CDM) halos
that are virialized clumps of CDM formed through gravitational 
instability from initial density perturbations in the early universe. 
In this scenario, the problem of galaxy formation can be 
divided roughly into two parts: (1) the formation of CDM 
halos in the cosmic density field, and (2) the accretion
and cooling of gas into dark matter halos to form stars
and the structures of luminous galaxies. The buildup of 
dark matter halos is now fairly well understood using both analytic 
models (e.g. Press \& Schechter 1974; Bond \etal 1991; Sheth, Mo \& 
Tormen 2001) and $N$-body simulations (e.g. Navarro, Frenk \& White 
1996; Jing \& Suto 2000; Moore et al. 1998; Zhao et al. 2003a, b; 
Wechsler et al. 2002). However, how gas cools and assembles 
to form the luminous part of a galaxy is still an open question.

 The pioneering work by Binney (1977), Rees \& Ostriker (1977) 
and Silk (1977) assumed that the gas in a dark matter halo 
is heated by shocks as it collapses into dark matter halos.
The heated gas then cools via radiative processes, and forms
stars in the centers of dark halos. This assumption has been 
combined with the halo population predicted by CDM 
models to make predictions for the properties of the 
galaxy population, such as the luminosity function, the colour 
distribution, etc (White \& Rees 1978; White \& Frenk 1991). 
With modern computational facilities and numerical codes, 
many of the processes in galaxy formation can be studied in more 
detail with numerical simulations (e.g., Katz \& Gunn 1991; 
Cen \& Ostriker 1992; Springel \& Hernquist 2003).

 Despite the progress made so far, galaxy formation remains 
one of the most challenging problems in astrophysics. 
One of the long-standing problems in the CDM scenario of galaxy 
formation is to match the halo mass function with the luminosity 
function of galaxies. The galaxy luminosity function 
has a characteristic shape that is very different  
from the halo mass function. The faint end slope of the 
galaxy luminosity function is significantly shallower than 
the halo mass function at the low-mass end. Also, the 
galaxy luminosity function has an abrupt break 
at the bright end, while such decline in the halo mass function 
occurs at a mass that corresponds to a much lower abundance (Yang \etal 2003). 
Furthermore, as shown in Yang \etal (2005), the relation 
between the luminosity of the central galaxy in a halo 
and the halo mass is strongly nonlinear at both the 
low-mass and high-mass ends. These results imply
that the efficiency of star formation must be suppressed 
in both low-mass and high-mass halos. 

Various suggestions have been made to explain the 
dependence of star formation efficiency on halo mass. 
For massive halos, it has been suggested that the low 
star formation efficiency is a result of the inefficient 
gas cooling. Thoul \& Weinberg (1995) studied gas cooling
in massive halos in considerable detail and found that 
a significant amount of gas can still cool in such halos, 
although the cooling efficiency is lower than that in 
lower mass halos. Thus, radiative cooling alone may not be 
able  to explain the bright-end behavior in the galaxy 
luminosity function. For low mass halos, suggestions have 
been made that heating by the UV background and/or 
supernova explosions associated with star formation may 
suppress gas cooling and hence the star formation efficiency.  
However, heating by the UV background is effective only 
in small halos with masses $M< 10^{10}\hMsun$ at low redshift 
(Quinn \etal 1996; Gnedin \etal 2000; Hoeft \etal 2006) and its 
effect is reduced by self-shielding at high redshifts 
(Dijkstra \etal 2004). Heating by supernova explosions 
may also be ineffective because of the development of 
Rayleigh-Taylor instabilities in the interstellar medium
(e.g. Mac Low \& Ferrara 1999; Strickland \& Stevens 2000).

One important process that can prevent gas from cooling too fast in dark 
matter halos is preheating (e.g. Mo \& Mao 2002; Oh \& Benson 2003; 
Scannapieco \& Oh 2004; Mo \etal 2005). In this scenario, 
the intergalactic gas is assumed to be heated to some finite entropy 
before it is accreted into dark matter halos. This will affect 
how gas is accreted into dark matter halos (e.g. Mo \& Mao 2002;
van den Bosch, Abel \& Hernquist 2003), and the distribution,   
thermal state and radiative cooling of the accreted gas.   
In this paper, we use 1-D simulations to study in detail the 
effects of preheating on galaxy formation in dark matter halos. 
We find a number of scaling relations that are useful to
understand and to model galaxy formation in preheated media.
Furthermore, we consider two realistic cases of preheating
and study their consequences on the formation of the central 
galaxies in dark matter halos. In one, preheating is assumed to 
be generated by the energy feedback of AGNs.
We show that such preheating can effectively suppress
gas cooling in halos more massive than $\sim 10^{13}\Msun$.
In the second case, we consider a constant, relatively low 
entropy floor and study its effect on mass accretion and cooling 
in low-mass halos. This model is motivated by the fact that
low-mass halos at low redshift may be surrounded 
by a warm intergalactic medium generated by gravitational 
pancaking (Mo \etal 2005). Our results show that the cold gas fraction 
in small dark matter halos can be significantly reduced in such a 
warm medium.

This paper is organized as follows. 
We briefly describe the numerical scheme of our 1-D hydrodynamical code 
and simulation models in 
Section\,\ref{sec:code}. In Section\,\ref{sec:adiabatic}, we show adiabatic 
simulation results and discuss the effects of preheating on gas accretion.
In Section\,\ref{sec:cooling}, we study the cooling of preheated media in 
halos of different masses using both single-phase and multi-phase simulations. 
We implement our simulations to study the two particular preheating models in 
Section\,\ref{sec:application}. 
Further discussion and a summary of our results 
are given in Section\,\ref{sec:sum}.

\section{Model and Simulation}
\label{sec:code}

In this paper, we use 1-Dimensional Lagrangian hydrodynamical simulations 
to study the accretion and cooling of gas in dark matter halos. 
In these simulations, the mass distribution is assumed to be spherically 
symmetric, and the radial distributions of the dark matter and gas 
are represented by spherical shells, each of which contains 
a certain amount of mass. The motions of the dark matter shells are 
followed with the same 1-d code as described in Lu \etal (2006). 
In what follows, we describe how we simulate the evolution of the
gas component.

\subsection{Hydrodynamics and thermodynamics}
We treat the gas component as fluid, so the motion is governed 
by the following set of equations:
\begin{eqnarray}
\frac{{\rm d}\rho_g}{{\rm d}t}+\rho_g \bigtriangledown \cdot v_g = 0\,; \\
\frac{{\rm d}v_g}{{\rm d}t}= -\frac{\bigtriangledown p}{\rho_g} - 
\bigtriangledown \Phi\,; \\
\frac{{\rm d}u}{{\rm d}t}=
      \frac{p}{\rho^2_g}\frac{{\rm d}\rho_g}{{\rm d}t} 
       + \frac{\Gamma-\Lambda}{\rho_g}\,; \\
p=(\gamma-1)\rho_g u\,.
\end{eqnarray}
These are the continuity equation, the momentum equation, 
the energy equation, and the equation of state, respectively. 
In these equations, $\rho_g$, $p$ and $u$ are, respectively, 
the density, pressure and specific thermal energy of the gas; 
$v_g$ is the fluid velocity; $\Phi$ is the gravitational potential; 
$\gamma$ is the adiabatic index (assumed to be $5/3$ in 
the following); $\Gamma$ and $\Lambda$ are the specific 
heating and cooling rates.  For spherical symmetry, the 
continuity equation and momentum equation can be re-written as
\begin{eqnarray}
{\rm d}m_g&=&4\pi r^2_g \rho_g{\rm d}r_g\,; \\
\frac{{\rm d}v_g}{{\rm d}t}&=&-4\pi r^2_g 
\frac{{\rm d}p}{{\rm d}m_g} - \frac{GM(<r_g)}{r_g^2}\,,
\end{eqnarray}
where $M(<r_g)$ is the total mass (including dark matter) 
enclosed by a sphere with radius $r_g$ and $G$ is gravitational constant. 
Note that the specific thermal energy of the gas can 
change not only by cooling and heating but also by 
adiabatic expansion and compression. Instead of the energy 
equation, one may choose to integrate the entropy equation. 
We define the specific entropy as 
\begin{equation}
S=\frac{T}{n^{2/3}}\,,
\end{equation}
where $T$ is the temperature, and $n$ is the particle number density 
of the gas. We can then write the equation of state as
\begin{equation}
p=A\rho_g^{\gamma},
\end{equation}
where the adiabat $A$ is related to $S$ as 
\begin{equation}
A=\frac{k}{\mu m_{\rm H}} \cdot S, 
\end{equation}
where $k$ is the Boltzmann constant, $\mu$ is the mean molecular weight 
and $m_{\rm H}$ is the mass of a hydrogen atom. 
The adiabat can be changed when the gas is heated or cools, 
or as it is shocked. We can therefore write 
\begin{eqnarray}
{\rm d}A&=&\frac{(\gamma-1) {\rm d}Q}{\rho_g^{\gamma-1}}\,; \\
{\rm d}Q&=&{\rm d}W_{\rm shock}+ {\rm d}Q_{\rm hc}\,,
\end{eqnarray}
where ${\rm d}W_{\rm shock}$ is the work done by shock,
and ${\rm d}Q_{\rm hc}$ is the heat due to other cooling and 
heating processes.

In most of the simulations presented in this paper, atomic radiative 
cooling is incorporated using the cooling 
function given in Katz, Weinberg \& Hernquist (1996).
We assume the gas has a primordial abundance, i.e.   
76\% of the mass is in hydrogen, 24\% is in helium, and there are no 
heavier elements. The atomic cooling processes considered here work 
only for temperature above $\sim 10^4$K. We calculate the 
abundance of different ionic species as a function of density and 
temperature assuming ionization equilibrium. Since the shock front 
is thin compared to all the length scales of interest, the mass of 
gas involved in the shock front is only a small fraction. 
Following Thoul and Weinberg (1995), we ignore cooling in 
the shock front. 

\subsection{Numerical scheme}

We use the Lagrangian finite-difference scheme of Bowers \& Wilson (1991)
to follow the evolution of gas shells. In this scheme, phase space 
quantities, such as positions and velocities are zone-edge-centered, 
while thermodynamic quantities, such as pressure, density and entropy/energy 
are zone-centered. In what follows, subscripts denote the index/position 
of gas shells. The acceleration of the $i$th gas shell is 
\begin{equation}
a_i=-4\pi r_i^2 \frac{p_{i+}-p_{i-}}{m_i} + g_i\,,
\end{equation}
where $p_{i+}$ and $p_{i-}$ are the pressures of the gas at the outer
and inner sides of the shell, $m_i$ is mass of the gas shell, 
and $g_i$ is gravitational acceleration.

The time integration is carried out by the DKD (drift-kick-drift) leap-frog 
algorithm:
\begin{eqnarray}
r_i^{n+1/2}&=&r_i^n + {1\over2}v_i^n {\rm d}\,t^n\,; \\ 
v_i^{n+1}&=&v_i^n + a_i^{n+1/2} {\rm d}\,t^n\,; \\
r_i^{n+1}&=&r_i^{n+1/2} + {1\over2}v_i^{n+1} {\rm d}\,t^n\,, 
\end{eqnarray}
in which superscripts denote the current time step. 
Once the integration is carried out to the $n$th step, we can calculate 
the gas density between every two neighboring shells: 
\begin{equation}
\rho_{i-}^{n}=\frac{3 m_{i}}{4\pi
\left[ (r_i^{n})^3-(r_{i-1}^{n})^3\right]}\,.
\end{equation}
During this time step, the change of adiabat is
\begin{equation}
\triangle A_{i-}^n=\frac{\gamma-1}
{(\rho_{i-}^n)^{\gamma-1}}(\triangle W_{{\rm shock},i-}^n
+ \triangle Q_{{\rm hc},i-}^n)\,,
\end{equation}
where the work done by the shock is given by
\begin{equation}
\triangle W_{{\rm shock},i-}^n = \frac{p_{{\rm shock},i-}^n \cdot
\triangle V_{i-}^n}{m_{i-}}=p_{{\rm shock},i-}^n \cdot
\left({1\over \rho_{i-}^{n+1}} -
{1\over \rho_{i-}^n}\right).
\end{equation}
Here $p_{\rm shock}$ is equivalent to the shock pressure using an 
artificial viscosity technique (Richtmyer \& Morton 1967; 
Thoul \& Weinberg 1995):
\begin{eqnarray}
p_{{\rm shock},i-}&=&-\frac{2\alpha}{1/\rho_{i-}^{n+1}
+1/\rho_{i-}^n} \cdot \vert v_i^{n+1/2}-v_{i-1}^{n+1/2} \vert
\nonumber \\
& &\times
(v_i^{n+1/2}-v_{i-1}^{n+1/2})\,,
\end{eqnarray}
if $v_i^{n+1/2}-v_{i-1}^{n+1/2}<0$, and $p_{{\rm shock},i-}=0$ otherwise.
Following Thoul \& Weinberg (1995), we choose $\alpha=4$.
Our tests show that the results are not sensitive to the value of this
parameter. The other entropy source term comes from radiative cooling, 
\begin{equation}
\triangle Q_{{\rm hc},i-}^n
= - \frac{\Lambda_{i-}^n}{\bar{\rho}_{i-}^n} \triangle t^{n+1/2}\,,
\end{equation}
where $\bar{\rho}_{i-}^n=\sqrt{\rho_{i-}^n \rho_{i-}^{n+1}}$ is the mean
density of the gas shells during the $n$th time step.
The advanced adiabat at the next time step is
\begin{equation}
A_{i-}^{n+1}=A_{i-}^n + \triangle A_{i-}^n\,.
\end{equation}
Other thermal quantities at the new time step are derived as follows:
\begin{eqnarray}
p_{i-}^{n+1} &=& A_{i-}^{n+1} (\rho_{i-}^{n+1})^{\gamma}\,; \\
u_{i-}^{n+1} &=& \frac{A_{i-}^{n+1}}{\gamma-1} (\rho_{i-}^{n+1})^{\gamma-1}\,.
\end{eqnarray}
We ignore radiative cooling on the shell front,
at which $v_i^{n+1/2}-v_{i-1}^{n+1/2}<0$ and $v_i\gg C_{s,i-}$, where
$C_{s,i-}=\sqrt{\gamma p_{i-}/\rho_{i-}}$ is the adiabatic sound speed
for the shell.

When a gas shell cools to a temperature below $10^4$K, 
the gas in the shell is considered to be cold. Since we do not include 
any cooling processes below this temperature, the cold gas is assumed to 
retain a temperature of $10^4$K until it flows into the center of 
the halo, where we drop the cold gas out of the gas component. 
Once a gas shell is dropped out, we do not trace its dynamical 
evolution anymore. Instead, we put the gas in a central exponential disk 
with a scale-length $r_{\rm d}={0.05\over \sqrt2} r_{\rm v}$ 
(Mo, Mao \& White 1998), where 
$r_{\rm v}$ is the virial radius of the halo at $z=0$. The cold gas 
disk is assumed to be a rigid object, and its gravity is included 
in the subsequent evolution of the other mass shells. 

Each time step of the integration is controlled to be
smaller than any of the important timescales:
the dynamical, the Courant, and the cooling timescales, 
\begin{eqnarray}
{\Delta}t_{\rm dyn} &=&\min_i\left\{c_d\sqrt{\frac{5r_i^3}{2GM_i}}\right\}\,, \\
{\Delta}t_{\rm Cour}&=&\min_i\left\{c_C\vert\frac{r_i-r_{i-1}}{C_s}\vert\right\}\,,\\
{\Delta}t_{\rm cool}&=&\min_i\left\{c_c\vert\frac{u_i\rho_i}{\Lambda_i}\vert\right\}\,,
\end{eqnarray}
where $c_d$, $c_C$ and $c_c$ are safety parameters. 
Moreover, in order to avoid shell crossing, we have an additional time step
control,
\begin{equation}
{\Delta}t_{\rm vel}=\min_i\left\{c_v\vert\frac{r_i - r_{i-1}}{v_i - v_{i-1}}
\vert\right\}\,,
\end{equation}
where $c_v$ is again a safety parameter. 
In practice, the time step is chosen to be the minimum of 
all the above time steps,
\begin{equation}
{\rm d}t=\min\{{\rm d}t_{\rm dyn},
{\rm d}t_{\rm Cour}, {\rm d}t_{\rm cool}, {\rm d}t_{\rm vel}\}.
\end{equation}
The safety parameters are taken to be $c_d=0.01, c_C=0.2, c_c=0.1$, and
$c_v=0.05$, as suggested by Thoul \& Weinberg (1995).

 The numerical code was tested against idealized problems with 
analytical solutions, such as the Sedov solution and the self-similarity 
models of Bertschinger (1983). The code also reproduces the results 
of Thoul \& Weinberg (1995).

\subsection{Initial conditions}

The simulation starts from an initial condition, which specifies the 
position, velocity and other properties of each shell at a high redshift, 
$z=z_i$. We set up our initial condition based on the mass accretion 
histories (MAHs) of dark matter halos. $N$-body simulations show that  
the mass accretion histories of CDM halos have some universal 
behaviors (e.g. Wechsler \etal 2002; Zhao \etal 2003a; b), and 
can be described by a simple functional form,
\begin{equation}
M(a)=M_0 \exp\left[-2 a_c \left({a_0\over a}-1\right)\right]\,,
\end{equation}
where $a$ is the expansion scale factor, $a_c$ is the scale factor 
corresponding to the formation time of the halo, 
and $M_0$ is mass of the halo at the observation time $a_0$
(Wechsler \etal 2002). Note that $a_c$ is the single free parameter.

Once the mass accretion history is specified, we can use it to 
determine the density profile of the initial perturbation.
In an expanding universe, a mass shell turns around and then 
collapses at a time that depends on the initial over-density 
within the mass shell. Based on a spherical collapse model, 
a mass shell collapses at a time when the linear over-density 
reaches $\approx 1.686$. Thus, the initial over-density within 
a mass  shell that collapses at a redshift $z$ can be written as 
\begin{equation}
\delta_i(M)= 1.686 \frac{D(z_i)}{D(z)},
\end{equation}
where $D(z)$ is the linear growth factor, and 
$M$ is the mass within the mass shell. The mass $M$
is related to the initial radius $r_i$ of the mass shell by
\begin{equation}
r_i=\left[\frac{M}{{4\over3} \pi \bar{\rho}(z_i) 
[1+\delta_i(M)]}\right]^{1/3}, 
\end{equation}
where $\bar{\rho}_i=\rho_{\rm crit,0} \cdot \Omega_M \cdot (1+z_i)^3$
is the mean matter density at redshift $z_i$, 
$\rho_{\rm crit,0}$ is the critical density of the universe at 
the present time, and $\Omega_M$ is the matter density parameter. 
For a given cosmology and halo mass accretion 
history, the above relations allow us to specify $\delta_i$ as a 
function $r_i$.

The initial velocity of each mass shell has two components, the Hubble 
expansion $v_i = r_i H(z_i)$, and the peculiar velocity owing 
to the density perturbation:
\begin{equation}
{\bf v} = \frac{2f}{3H \Omega_M}{\bf g}\,,
\end{equation}
where $\bf g$ is the peculiar gravitational acceleration,
\begin{equation}
{\bf g}({\bf x})=G\rho_b a \int {\rm d}^3 x' 
\delta({\bf x}',t)({\bf x-x'})/\vert {\bf x'-x}\vert^3\,, 
\end{equation}
and $f={\rmd}\ln D(z)/\rmd\ln{(1+z)}$, with all the quantities evaluated 
at the initial redshift $z_i$. We choose $z_i=200$, and the initial 
temperature of gas shells is set to be the CMB temperature at this epoch.

Our simulations cover a large range of halo mass, from $10^{10}$ 
to $10^{15}\hMsun$. In hierarchical structure formation, the value of $a_c$ 
is correlated with halo mass. In our simulations, we use the typical 
value of $a_c$ for halos of a given mass. The parameters of the halo 
models are described in Table~\ref{tab:halo}. As in Lu et al. (2006), 
particles accreted before $z_c$ are assumed to have isotropic 
velocities, so that the final dark halos have the 
Navarro, Frenk \& White profiles (Navarro et al. 1996, 1997).

\begin{table*}
\begin{center}
\caption{Halo models}\label{tab:halo}
\begin{tabular}{lcccccc}
\hline\\
M ($\hMsun$)& $10^{10}$ & $10^{11}$ & $10^{12}$ & $10^{13}$ & $10^{14}$ & $10^{15}$ \\
\hline \\
$a_c$   & 0.2   & 0.23    & 0.3     & 0.4     & 0.6     & 0.8 \\
$z_c$   & 4.0   & 3.3     & 2.3     & 1.5     & 0.67    & 0.25 \\
\hline
\end{tabular}
\end{center}
\end{table*}

In all the simulations, we assume $\Omega_M=0.3$ 
and $\Omega_{\Lambda}=0.7$ at the present time. 
We fix the baryon fraction to be $f_b=0.17$ 
for our fiducial model, but vary this parameter in some of our simulations.
We use $5\times10^4$ shells of the same mass for the dark matter,  
and 500 equal-mass shells for the gas component. To avoid a numerical 
instability caused by dark matter shell-crossing, the mass of each dark matter 
shell is chosen to be smaller than that of the gas shell. 
Our tests with higher resolutions show that 
our resolution ensures numerical convergence. 

\section{Accretion by dark matter halos in a preheated medium} 
\label{sec:adiabatic}

\begin{figure*}
\centerline{\psfig{figure=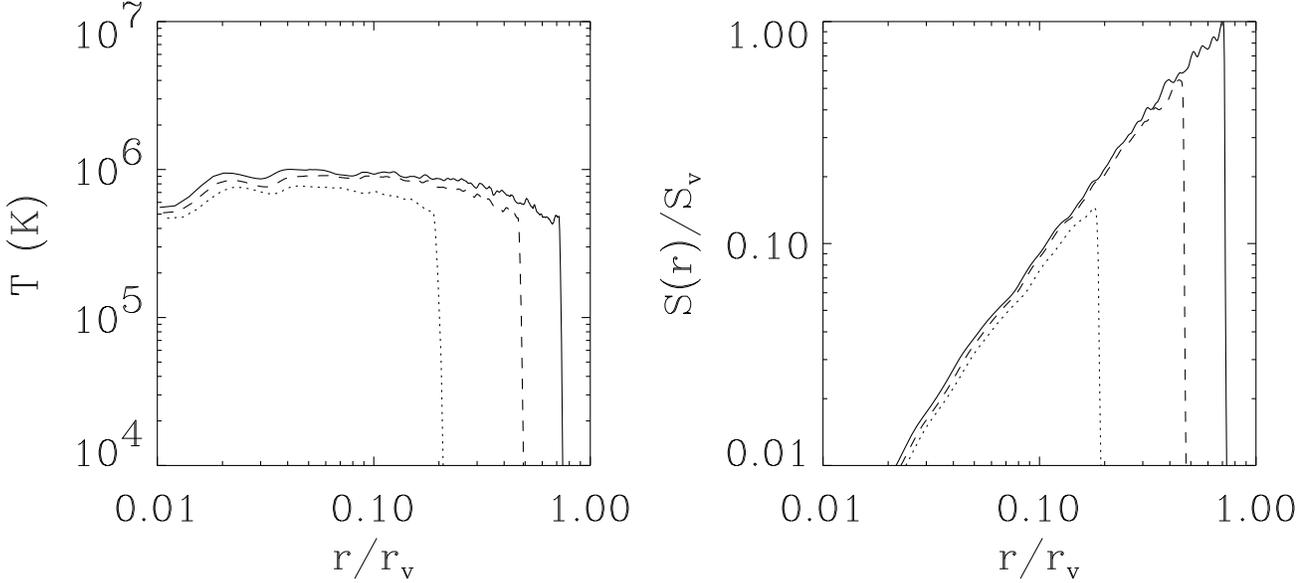,width=\hdsize}}
\caption {
The evolution of temperature (left panel) and normalized entropy 
profile (right panel) of a $M=10^{12}\hMsun$ halo in the absence of 
preheating. The dotted, dashed, and solid curves correspond 
to $z=3$, $1.5$ and $0.85$, respectively. 
The radius is scaled to the virial radius of the halo 
at $z=0$. The sharp breaks correspond to the shock fronts.
}
\label{fig:temp_entropy}
\end{figure*}

In this section, we present simulation results that neglect radiative 
cooling. First, we consider models in which the 
intergalactic medium is cold. During the formation of a dark matter halo, 
gas is accreted and shocked. In the absence of radiative cooling, 
the temperature of the post-shock gas is roughly given by the in-fall 
kinetic energy of the gas:
\begin{equation}
T_{ps}=\frac{3\mu m_H v_s^2}{16k},
\end{equation}
where the shock velocity $v_s$ is approximately the virial velocity 
of the halo at the time in consideration. Since the gravitational 
potential of the halo is established at an early stage of halo 
formation, the virial velocity does not change much at later times.
This is the main reason for the rather flat 
temperature profiles shown in the left panel of 
Figure\,\ref{fig:temp_entropy}. Note that, as the halo mass increases with time, 
the radius within which the gas is shock heated also increases.

\begin{figure}
\centerline{\psfig{figure=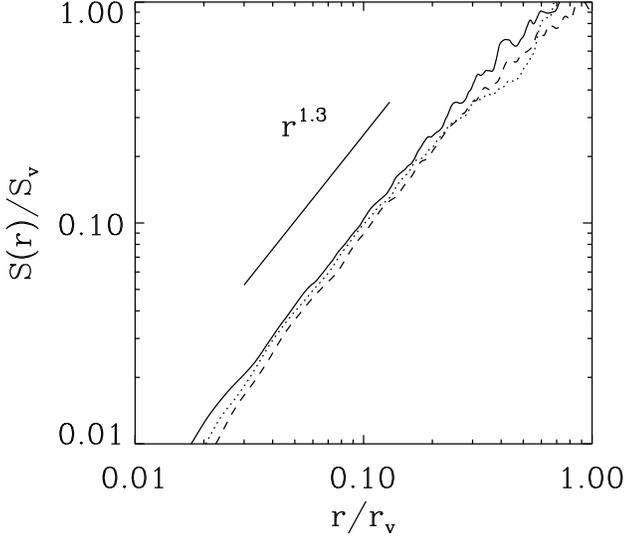,width=\hssize}}
\caption {The normalized specific entropy profiles for 
halos of different masses at $z=0$. 
The solid, dashed and dotted curves are for
halos with masses of $10^{10}$, $10^{12}$ and $10^{14}\hMsun$, 
respectively. We scale the radius to the 
virial radius, $r_{\rm v}$, and scale the specific entropy 
to the virial entropy, $S_{\rm v}$, 
as defined in the text.}
\label{fig:s_r_0}
\end{figure}

The right panel of Figure\,\ref{fig:temp_entropy} shows the evolution 
of the specific entropy profile. The specific entropy of the accreted 
gas increases with radius, because the density of shocked gas is lower 
at larger radius. The entropy profile is roughly a power-law and shows
a self-similar behavior. Indeed, if scales the radius $r$ in units of the 
present virial radius, and scales the specific entropy in units of 
the ``virial entropy'', defined as
\begin{eqnarray}
\sv=\frac{T_{\rm v}}{n_{\rm v}^{2/3}}\,,
\end{eqnarray}
where
$n_{\rm v}$ is the mean particle density within the virialized 
halo, the entropy profiles for halos of different masses 
all obey:

\begin{equation}
\frac{S(r)}{\sv}=\left(\frac{r}{r_{\rm v}}\right)^{\alpha},
\end{equation}
with $\alpha\sim 1.3$, as shown in Figure\,\ref{fig:s_r_0}. This 
result is consistent with the analytical work of Tozzi \& Norman 
(2001), but the slope is steeper than the observed entropy profiles 
of nearby cooling flow clusters (Piffaretti etal 2005), which 
have $\alpha\sim 0.95$. As we shall see, radiative cooling 
changes the entropy profile. 

\begin{figure}
\centerline{\psfig{figure=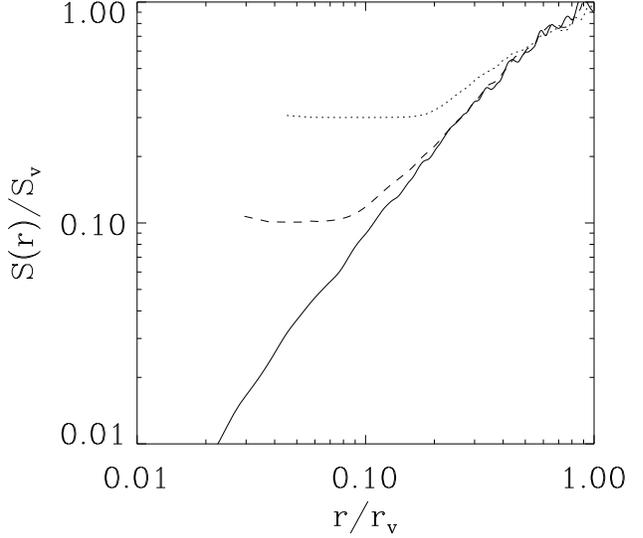,width=\hssize}}
\caption 
{The entropy profiles in two preheating models:
the dotted line corresponds to $\sph=0.3\sv$ and the dashed line 
corresponds to $\sph=0.1\sv$, compared to 
the model without preheating (solid line). 
We ignore radiative cooling.
}
\label{fig:s_r_ph}
\end{figure}

 Next, we consider cases where the gas is preheated to a finite entropy
before it is accreted into dark matter halos. In the simulations, the 
preheating entropy is gradually added into gas shells at early epochs,  
from $z=100$ to 10, to make the numerical calculations stable. 
Here again 
we neglect radiative cooling. In Figure\,\ref{fig:s_r_ph} we show the entropy profiles
for cases with different initial entropies. One can clearly see
the imprint of the initial entropy. The entropy profile is flat 
at small radii, with an amplitude comparable to the initial 
entropy, since the entropy generated by accretion
shocks is smaller than the initial entropy at small radii. Since the entropy 
generated by shocks increases with radius, 
the entropy profile will eventually be dominated by shock generation 
and follow that of no preheating case. This transition occurs roughly at 
the radius where the shock entropy equals to the preheating entropy. 
The final entropy profile can, therefore, roughly be described by
\begin{equation}
S(r)= \left\{ \begin{array}{lll}
S_{\rm ph} & \mbox{if} & r<r_t; \\
S^0(r)    & \mbox{if} & r \geq r_t, \end{array} \right.
\end{equation}
where $S_{\rm ph}$ is the preheating entropy, $S^0(r)$ is the
entropy expected from cold accretion, and $r_t$ is the radius 
at which $S^0=S_{\rm ph}$. Since $S^0(r)/\sv$ is roughly a scale-free 
function of $r/r_{\rm v}$, the characteristic radius $r_t$, expressed in units of 
$r_{\rm v}$, is determined by $S_{\rm ph}/S_{\rm v}$.
Hence, the effect of preheating in adiabatic accretion is almost
entirely determined by the ratio $\sph/\sv$. 

\begin{figure}
\centerline{\psfig{figure=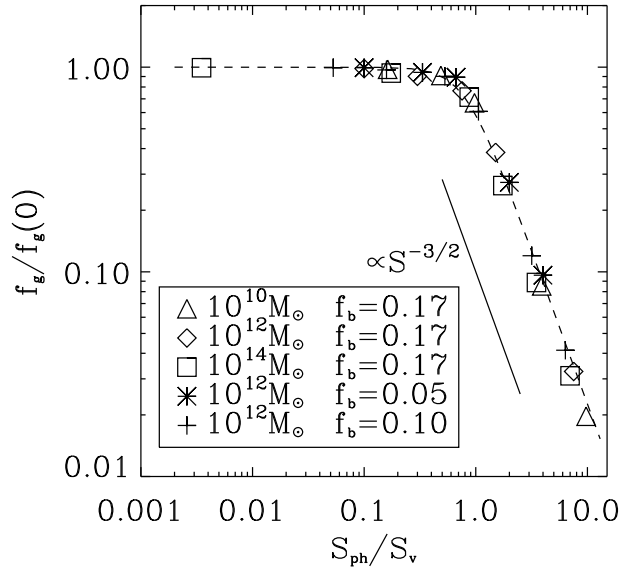,width=\hssize}}
\caption {
The gas fraction that can be accreted into a halo
in the presence of preheating, normalized by the fraction 
in the absence of preheating, as a function of 
normalized preheating entropy, $S_{\rm ph}/S_{\rm v}$. Different symbols 
represent simulations with various halo masses and baryon fractions, as 
indicated in the panel. The dashed curve corresponds to 
equation (\ref{eq_fgf0}), while the solid line corresponds 
to the scaling relation, $\propto S_{\rm ph}^{-3/2}$.
}
\label{fig:fclps}
\end{figure}

As expected, preheating reduces the amount of gas that 
accretes into a halo. In Figure\,\ref{fig:fclps} we show the ratio of 
$f_g(\sph)/f_g(0)$ as a function of $\sph/\sv$, where 
$f_g(\sph)$ is the ratio between the mass of accreted gas and 
the halo mass, and $f_g(0)$ is the corresponding ratio for the non-preheating
model. As one can see, the $f_g(\sph)/f_g(0)$ - $\sph/\sv$ relation 
is independent of both the halo mass and of the initial baryonic fraction.  
The relation can be well described by
\begin{equation}\label{eq_fgf0}
f_g/f_g(0)=
\frac{1}{\left[1+\left({\sph/\sv \over 0.8}\right)^3\right]^{1/2}}\,.
\end{equation}
This relation shows that the reduction in the accreted gas
becomes important when the preheating entropy is comparable 
to or larger than $\sv$. For $\sph \gg \sv$, 
$f_g/f_g(0)\propto (\sph/\sv)^{-3/2}$, and if $\sph= {\rm constant}$ 
then $f_g/f_g(0)$ is proportional to the halo mass, $M$. In  
the last relation we assumed the virial density $n_{\rm v}$ is  
independent of the halo mass. This relation is similar to 
that obtained by Mo \& Mao (2002) based on a simple analytical
model.  
 
\section{Cooling of preheated gas in dark matter halos}
\label{sec:cooling}

Since preheating modifies the distribution of the accreted gas 
in dark halos, it can change the gas cooling rate 
(Mo \& Mao 2002). In this section, we explore how preheating 
affects gas cooling in different halos. 

\subsection{Cooling in single-phase media}

\begin{figure*}
\centerline{\psfig{figure=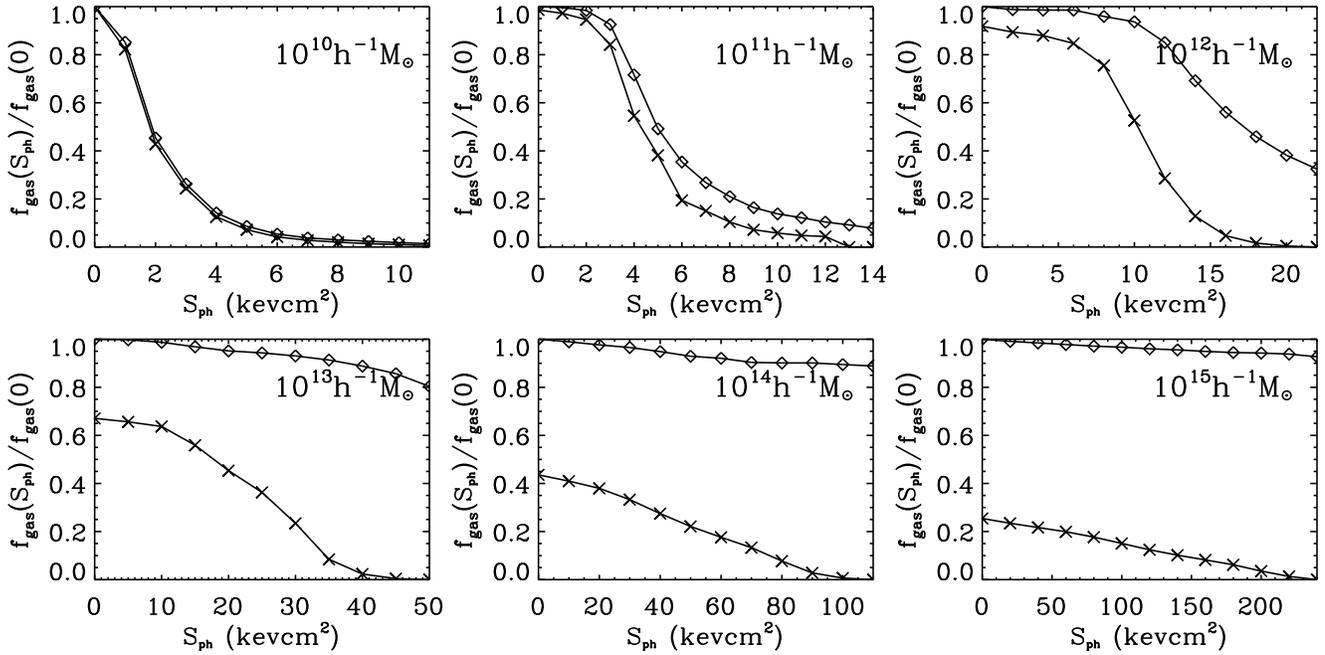,width=\hdsize}}
\caption {Diamonds show the amount of accreted gas in the 
presence of preheating, normalized by the amount of 
accreted gas in the absence of preheating, 
versus the preheating entropy.   
Crosses show the amount of cold gas, normalized by the 
amount of accreted gas in the absence of preheating,
versus the preheating entropy.     
Different panels show results for halos of different masses.
}
\label{fig:fcc}
\end{figure*}

 To begin, we assume that the accreted gas in a dark halo 
has a single-phase. Thus, at any given time, each gas shell
has a unique temperature and density.
In the simulation, the accreted gas dissipates its thermal energy 
via radiative cooling, and its temperature drops. We consider the gas 
to be in a cold component when its temperature 
drops below $10^4$K. To characterize the efficiency of cooling, 
we study the fraction of the accreted gas that is in the cold 
component at $z=0$. For simulations with preheating, we denote the 
total cold gas mass and the total accreted gas mass by
$M_{\rm cool}(\sph)$ and $M_{\rm acc}(\sph)$, respectively. 
The corresponding quantities in the non-preheating cases
are denoted by $M_{\rm cool}(0)$ and $M_{\rm acc}(0)$.
Figure\,\ref{fig:fcc} shows the ratio 
$f_{\rm cool}(\sph)\equiv M_{\rm cool}(\sph)/M_{\rm acc}(0)$ and
$f_{\rm acc}(\sph)\equiv M_{\rm acc}(\sph)/M_{\rm acc}(0)$
as a function of the preheating entropy, $\sph$, for halos of different 
masses. As one can see, the behavior of $f_{\rm cool}(\sph)$ 
is quite different for different mass halos. 
In massive halos, the accreted gas fraction decreases slowly
with increasing amount of preheating entropy, while the cold gas fraction drops 
rapidly. Thus, in these systems, a large amount of the accreted gas 
stays in the hot phase. The situation is quite 
different for low mass halos, where almost all of the 
accreted gas can cool even if the IGM is preheated. 

\begin{figure}
\centerline{\psfig{figure=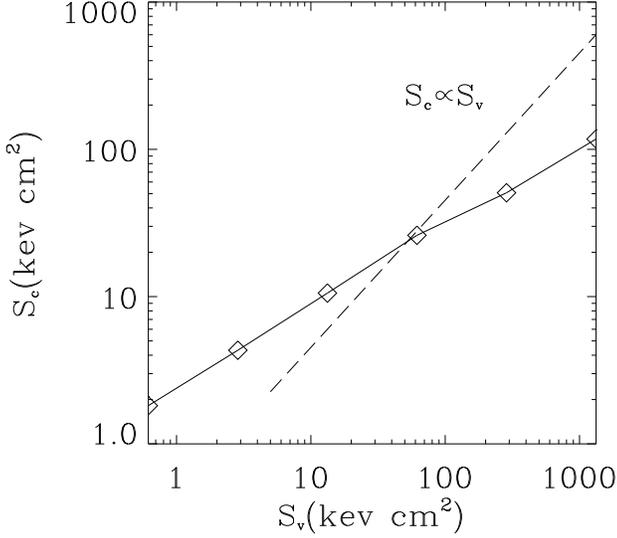,width=\hssize}}
\caption {The critical entropy $S_{\rm c}$, as defined in the text, 
versus halo virial entropy $\sv$. The dashed line shows 
$S_{\rm c}\propto\sv$.
}
\label{fig:sc}
\end{figure}

 To quantify the importance of preheating on gas 
cooling, we define a ``critical'' specific entropy $S_{\rm c}$ by
\begin{equation}
f_{\rm cool}(S_{\rm c})/f_{\rm cool}(0)=1/2\,,
\end{equation}
where $f_{\rm cool}(S)$ is the fraction of the gas that can cool
in a halo if the IGM is preheated to a specific entropy $S$. 
Figure\,\ref{fig:sc} shows the $S_{\rm c}-\sv$ relation obtained from our simulations.
Clearly, one needs a higher initial entropy to prevent the 
accreted gas from cooling in more massive systems.
However, the critical specific entropy, $S_{\rm c}$, does not increase 
as fast as $\sv$: the logarithmic slope of the $S_{\rm c}-\sv$ relation 
is approximately $2/3$ not 1. Thus, if there were a mechanism
that could preheat the gas associated with halos to an initial 
specific entropy that was proportional to $\sv$, the effect 
on gas cooling would be more important for more massive systems.
On the other hand, if the IGM were everywhere preheated to the same 
level of specific entropy, then the effect would be more important for 
lower-mass systems. 

\begin{figure}
\centerline{\psfig{figure=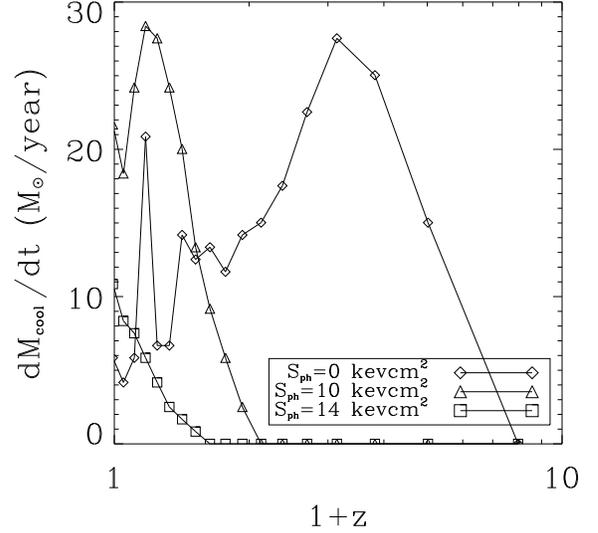,width=\hssize}}
\caption {The gas cooling rates in a $10^{12}\hMsun$ halo versus
redshift. Different symbols represent results for different preheating 
levels, as indicated in the panel.
}
\label{fig:chsingle}
\end{figure}

 It is also interesting to look at how preheating affects the gas 
cooling rate as a function of time. Since preheating flattens the 
gas density distribution, the time scale for gas cooling is 
similar over a large range of radii. This is different from 
the case without preheating, where the cooling time scale 
is shorter in the inner, denser regions and the cooling proceeds
in an inside-out fashion. In Figure\,\ref{fig:chsingle}, we show 
the cooling rate as a function of redshift in a halo of 
$10^{12}\hMsun$. Here we consider two models of preheating, 
one with $\sph=10\kevcm2$ and the other with $\sph=14\kevcm2$, 
and compare the results of these two models to the 
case with $\sph=0$. Againi, we assume the accreted gas has 
a single phase. As one can see, preheating causes a delay in 
gas cooling. The cooling rate 
peaks at $z\sim 2$ in the absence of preheating, and  
the cooling rate at the present time is about $4{\rm \Msun/year}$. 
Preheating decreases the cooling rate at high $z$ but enhances
the cooling rate at lower $z$, because much of the gas has to 
wait until a later time before it can cool.   
For all the simulations we analyzed, the delay in gas 
cooling is more important if the ratio $\sph/\sv$ is larger.

 Based on the results presented above, one can conclude that
the accretion of cold gas into galaxies occurs rather late in the 
presence of preheating. This may be a problem, because it implies 
that most of the stars in present spiral galaxies must form
very late. As we will see below, this problem can be alleviated
by assuming that the gas is in a multi-phase rather than in a single-phase  
state. 

\subsection{Cooling in multi-phase media}
\label{sec:multi}

 There are many reasons to believe that the gas in galaxy halos 
has a multi-phase structure. First of all, at the temperatures 
considered here, the gas is expected to be thermally unstable, 
and so a multi-phase medium can develop through thermal 
instability (Field 1965; Fall \& Rees 1985; Murray \& Lin 1990). 
Secondly, the density field of the CDM universe is expected to be clumpy, 
which can promote thermal instability. The development of a multi-phase 
medium will affect the cooling of the halo gas, as the cooling time in the 
high-density/low-temperature phase may become shorter than the age, 
even if the single-phase cooling time does not.     
In this section, we use simple models to demonstrate how gas cools in 
a multi-phase medium.

 To incorporate a multi-phase medium in our simulations,   
we split each gas shell into $N$ phases over a range of gas 
densities. The gas in each phase is assumed to move 
with its host shell and to be in pressure equilibrium with the other 
phases in any subsequent evolution. Each sub-shell, which represents 
gas in a phase, is characterized by its number density $n_{ij}$, 
temperature $T_{ij}$ and mass $m_{ij}$, where $i$ labels a shell, 
and $j$ labels a sub-shell phase. The thermodynamics of the 
multi-phase gas is treated in the same way as that described in 
Thomas (1988). At each time step, the radiative cooling of every 
sub-shell is calculated using its current density and temperature,
and a new density and a new temperature are determined 
through the hydrodynamical equations (see Thomas 1988 and its appendix
for details). Once the temperature of a sub-shell drops below 
$10^4$K, this phase drops out of the shell, and  
the gas in the sub-shell is then deposited into a central disk that 
is assumed to form at the center of the host halo. The remaining 
phases in the shell are assumed to expand adiabatically to occupy the extra volume. 
In practice, we adopt the multi-phase model for a gas shell as soon 
as it is accreted within the current virial radius. We, therefore, 
neglect the details of how a multi-phase medium actually develops. 

\begin{figure}
\centerline{\psfig{figure=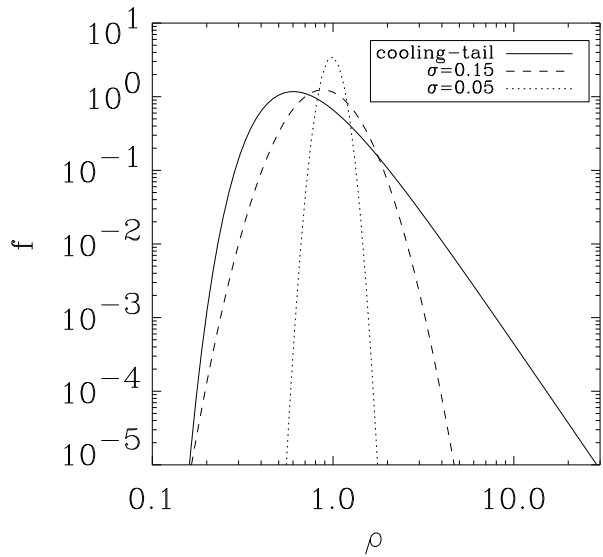,width=\hssize}}
\caption {
The volume fraction occupied by phases of different 
density. The dotted and dashed curves are log-normal distributions 
with dispersions $\sigma_{\log{\rho/\rho_0}}=0.05$ and $0.15$, 
respectively. The solid curve is the ``cooling tail'' model
described in the text. The x-axis is scaled by the mean 
density, $\bar{\rho}$.
}
\label{fig:phasedist}
\end{figure}

The density distribution of the multi-phase gas is described by the 
volume fraction function, $f(\rho, \bar{\rho}){\rm d}\rho$, which describes 
the volume fraction occupied by a phase with a density in the range 
$\rho\pm {\rm d}\rho/2$ for a gas shell with mean density $\bar{\rho}$. 
By definition $\int f\,{\rm d}\rho =1 $. In our simulations, each 
gas shell initially contains 50 sub-shells equally spaced in 
$\log \rho$. Since the form of $f(\rho, \bar{\rho})$ is not known
{\it a priori}, we consider three simple models, as examples, which are 
plotted in Figure\,\ref{fig:phasedist}. 
Two of these are log-normal functions with two different widths, 
$\sigma_{\log{ \rho/\rho_0}} = 0.15$ and $0.05$, where 
$\rho_0$ is the median density of the distribution, related to the mean density, 
$\bar{\rho}$, by $\rho_0=\bar{\rho}e^{-\sigma^2/2}$. 
In the third model, the volume-fraction function is assumed to
have the ``cooling tail" form, 
\begin{equation}
f(\rho, \bar{\rho}) \propto 
\rho^{-(4-\alpha)}\exp\left[-(\bar{\rho}/\rho)^{2-\alpha}\right]\,,
\end{equation}
where $\alpha$ is the slope of the cooling function, 
$\Lambda \propto T^{\alpha}$. 
This model has been used  
to describe multi-phase cooling flows in galaxy clusters
(Nulsen 1986; Thomas 1988; Waxman \& Miralda-Escud\'e 1995).
We have adopted $\alpha=0.4$, which roughly describes  
the cooling by free-free process. In the simulations, we only 
sample phases with $f>10^{-5}$. Our test with the ``cooling tail'' model 
shows that our simulation reproduces the results obtained 
by Thomas (1988).

\begin{figure}
\centerline{\psfig{figure=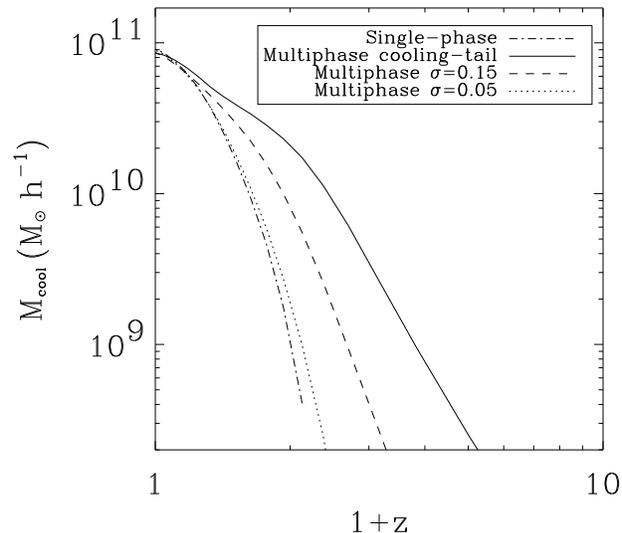,width=\hssize}}
\caption {The total amount of cooled gas in a $10^{12} \hMsun$ halo
with preheating entropy $S_{\rm ph}=S_{\rm c} \sim 10\kevcm2$. 
The dash-dotted line shows the result of a model that assumes the halo gas 
is a single-phase medium. The three other curves show the results 
of the three multi-phase models: the cooling-tail model (solid curve), 
the log-normal model with $\sigma=0.15$ (dashed curve), 
and log-normal model with $\sigma=0.05$ (dotted curve).
}
\label{fig:coolhist}
\end{figure}

To demonstrate the effect of changing the gas from a single-phase
medium to a multi-phase medium, we consider a case where the 
final halo mass is $10^{12} \hMsun$, and the initial preheating 
entropy $S_{\rm ph}=10\kevcm2$, using the three models for 
$f$ described above. Figure\,\ref{fig:coolhist} shows the
gas cooling histories, i.e. the mass of cooled gas as a function of 
redshift, compared to that of a 
single-phase model. As expected, the cooling history is more extended 
in models where the density distribution is broader. 
For a broader density distribution, there is larger amount of 
high-density gas that can cool rapidly. If the density distribution 
is as broad as the ``cooling-tail'' model, cold gas can be assembled 
into galaxies much earlier than in the single-phase model.

\begin{figure}
\centerline{\psfig{figure=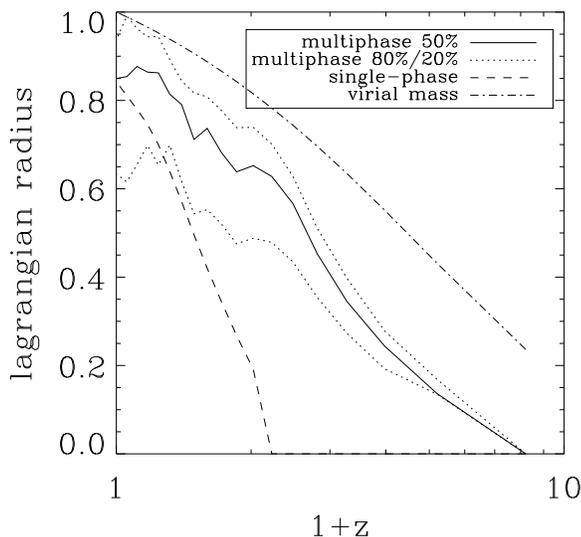,width=\hssize}}
\caption {
The Lagrangian radius of the gas shell that cools as a 
function of redshift normalized to the 
present time ($z=0$). Results are shown for a $10^{12}\hMsun$ halo 
with preheating entropy $\sph=S_{\rm c}=10\kevcm2$, and 
the ``cooling-tail'' model is adopted for the multi-phase medium. 
The solid curve shows the Lagrangian radius within which 
$50\%$ of the cooled gas is contained at each redshift, while
the two dotted curves mark the radii which contain 
20\% and 80\% of the cooled gas.
For comparison, the dashed curve shows 
a model in which the medium is assumed to be single-phase.
The dot-dashed curve is the Lagrangian radius of the forming halo.
}
\label{fig:cooldist}
\end{figure}

 Remember that in the single-phase model, cooling is inside-out. So 
if the cooling time of the gas were comparable to the age, only gas in 
the innermost region could cool. On the other hand, in a 
multi-phase medium, cooling can occur almost everywhere in 
the high-density phases. To demonstrate this, we plot in 
Figure\,\ref{fig:cooldist} the initial (Lagrangian) radii
that contain 20\%, 50\% and 80\% of the gas that drops out 
to the cold phase in a time interval $\Delta t= 0.5\,{\rm Gyrs}$ 
as a function of $(1+z)$. In the single-phase case, all the gas within 
a certain Lagrangian radius can cool, and so the 20\%, 50\% and 80\%
radii are all the same (the dashed curve in  Figure\,\ref{fig:cooldist}).
For the single-phase case shown in Figure\,\ref{fig:cooldist}, this 
``cooling radius'' is much smaller than the virial radius of the halo 
(dot-dashed curve), implying that only gas in the innermost parts 
of the halo can cool. In the multi-phase 
case, however, at any time gas can cool to the cold phase over a large 
range of Lagrangian radii. This result could have important 
implications for the formation of galaxy disks. $N$-body simulations 
show that the specific angular momentum of dark matter particles 
increases roughly linearly with the Lagrangian radius (Bullock 2001). Thus, if
the angular momentum of the gas follows that of dark matter,  
the specific angular momentum of the cooled gas in the single-phase
case is expected to be lower than the average specific angular 
momentum of the dark matter, since only the gas in the 
inner regions can cool. This could lead to the formation of 
galaxy disks that are too compact (e.g. Mo, Mao \& White 1998). 
On the other hand, in the multi-phase model, the cold gas that 
settles into the disk comes from various radii, and so its
average specific angular momentum may be comparable to that of the
dark matter, a condition that is required to reproduce 
the observed sizes of galaxy disks (e.g. Mo, Mao \& White 1998).

\begin{figure*}
\centerline{\psfig{figure=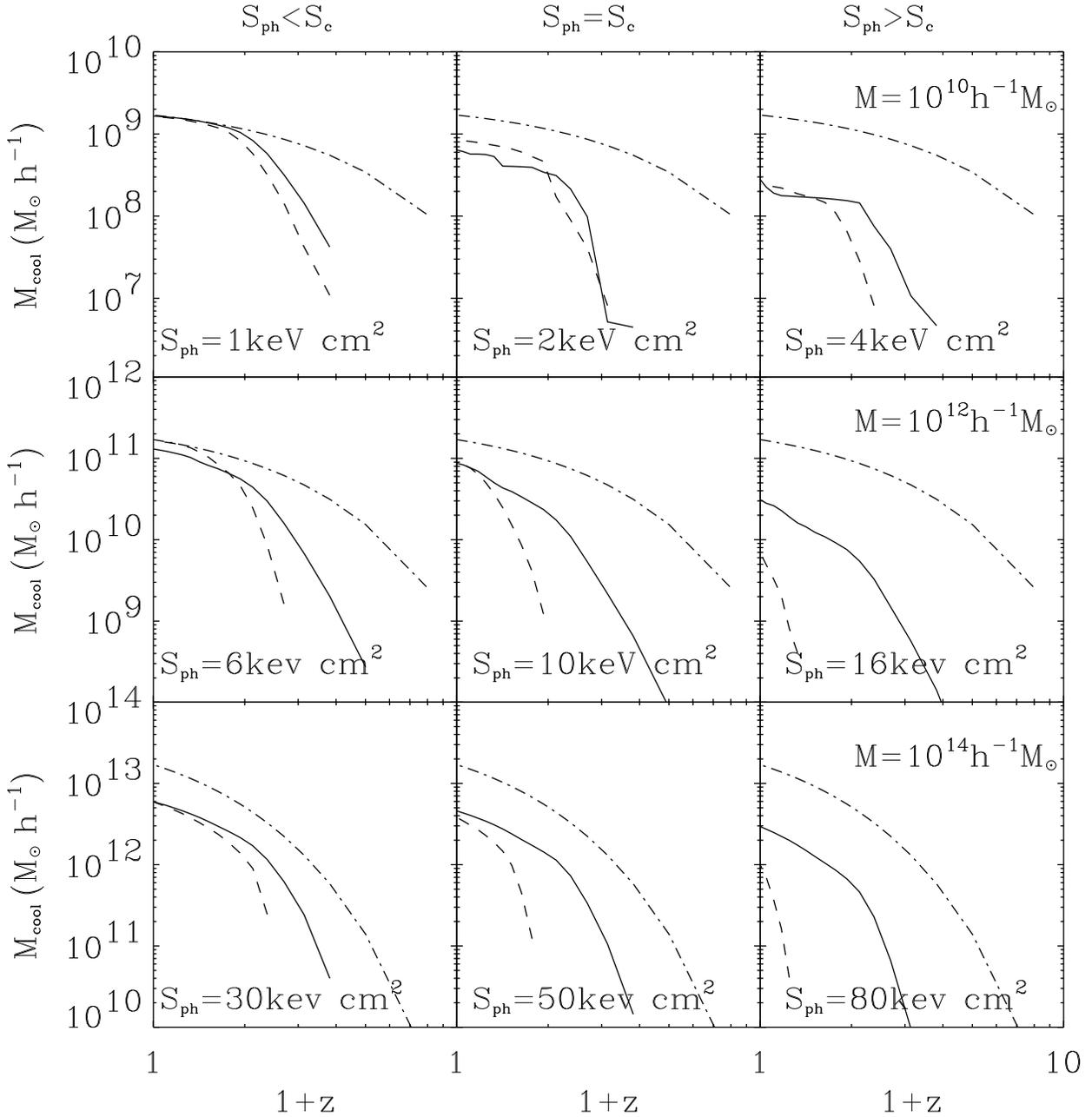,width=\hdsize}}
\caption {The mass of cooled gas as a function of redshift. 
In each panel, the solid curve shows cooled gas mass in  
a simulation assuming a multi-phase medium with the 
``cooling tail'' density distribution, and the dashed 
curve stands for the result of the corresponding single-phase 
simulation. The dash-dotted curve shows the baryon 
fraction ($f_b=0.17$) times the halo mass.
Results are shown for halos of different present masses which are
noted in the right column panels for each row
and different preheating entropies which are noted at the bottom 
of each panel.
}
\label{fig:ms}
\end{figure*}

As demonstrated in Figure\,\ref{fig:ms}, if the initial specific 
entropy, $S_{\rm ph}$, is comparable to or smaller than the critical 
value, $S_{\rm c}$, the total amount of gas that can cool at the present 
time is quite insensitive to the assumed gas density distribution, 
although different models of the gas density distribution
produce different accretion histories of cold gas.  
This occurs because the cooling time of the gas accreted into halos, 
even when it is assumed to be in a single-phase, is comparable to or shorter than 
the age of the universe at the present time. However, if the value of 
$S_{\rm ph}$ is much larger than $S_{\rm c}$, the total amount of gas that 
cools by the present time can be significantly larger in the 
multi-phase model than in the single-phase model, for halos with masses 
$\ga 10^{11}h^{-1}{\rm M}_\odot$. For halos with smaller masses, 
cooling is so effective that almost all the accreted gas can cool even 
if $S_{\rm ph}>S_{\rm c}$.

\section{Applications}
\label{sec:application}

So far we have discussed in general how preheating affects 
gas accretion and cooling in dark matter halos. 
In the present section, we consider two specific examples that may be 
relevant to the actual formation of galaxies.    

\subsection{Preheating by a central AGN}

First, we consider the case where the gas associated with a dark 
halo is heated by the outflow of a central AGN. Recent observations
show that almost all galaxies, at least those with
a significant spheroidal component, i.e. elliptical galaxies and 
spiral galaxies with a significant bulge, host a central supermassive 
black hole (SBH) with a mass that is correlated with the velocity 
dispersion of the bulge,
$M_{\rm BH}=10^{8.3}\Msun(\sigma/200{\rm km\,s^{-1}})^{4.02}$
(Tremaine etal 2002). This correlation is quite independent of 
redshift out to $z\approx 3$ (Sheilds \etal 2003).
If fed with gas, these black holes are able to power AGNs that may 
produce outflows and heat the gas in dark matter halos. 
This kind of feedback process has been suggested as a
plausible mechanism to quench both star formation and 
black halo growth in massive galaxies (Binney \& Tabor 1995; 
Silk \& Rees 1998; Nulsen \& Fabian 2000; Springel \etal 2005). 
However, it is still unclear how the energy associated with the AGNs
is coupled to the gas, allowing it to quench gas cooling and star formation.  
It has been suggested that the energy deposition is localized
so that the outflows driven by AGNs are thermalized in 
the ambient medium only within the host galaxy (e.g. Springel \etal 2005). 
However, recent numerical simulations with jets show that 
the jet may tunnel through low-density regions
of the ambient medium, and that the energy may be easily carried 
to large distances and thermalized there (Vernaleo \& Reynolds 2005). 
Without going into details, we consider a simple model here.
 
We assume that the energy injected by an AGN is proportional to the mass of 
the SBH, $E_{\rm AGN}\propto M_{\rm BH}$. Observations of spiral galaxies 
with SBHs show that the mass of the SBH is tightly correlated 
with the circular velocity of the galaxy, $M_{\rm BH}\propto v_{\rm c}^5$
(Ferrarese 2002). Since $v_{\rm c}$ is related to the virial mass of the 
spiral galaxy's host halo, $M\propto v_{\rm c}^3$, we have 
$E_{\rm AGN}/M\propto v_{\rm c}^2$. Thus, if the energy of the 
AGN is to heat all the gas associated with its host halo, the 
energy gained by each particle from the AGN feedback is roughly 
proportional to $v_{\rm c}^2$, i.e. to the virial temperature of the 
host halo. If such feedback occurs at the same redshift for 
systems with different masses, the specific entropy owing to AGN preheating
is expected to be proportional to the virial entropy of the
halo, i.e. $\sph \propto \sv$ (Scannapieco \& Oh 2004).
However, observations of quasars suggest that there is an upper limit 
to quasar luminosities (Fan \etal 2003), corresponding to a 
SMB mass of $M_{\rm BH}\sim 10^9\Msun$. This mass corresponds to
$v_{\rm c}=500 {\rm km\,s^{-1}}$ assuming Eddington luminosities and the observed 
$M_{\rm BH}-v_{\rm c}$ relation (Wyithe \& Loeb 2003). 
Such a limit in $M_{\rm BH}$ implies that the 
$M_{\rm BH}$-$v_{\rm c}$ relation must be broken for massive halos.
There must be some process that quenches the growth of the 
SMB and, given the $M_{\rm BH}$-$\sigma$ relation, also the 
growth of the central galaxy in massive halos. In what follows, 
we will consider a preheating model with $\sph \propto \sv$, 
and examine if the observed mass limit can be accommodated. 

Based on the above discussion, we write
\begin{equation}\label{AGN_sph}
\sph=S_{\rm ph,12}{\sv \over S_{\rm v,12}},
\end{equation}
where $S_{\rm v,12}$ is the virial entropy of a $10^{12}\hMsun$ halo and 
$S_{\rm ph,12}$ is a normalization factor, taken to be 
the value of $S_{\rm ph}$ corresponding to that in a halo
of present mass $M=10^{12}\hMsun$. To estimate 
the normalization factor, we use the model presented in 
Scannapieco \& Oh (2004). Since the lifetime of an AGN is typically 
$\sim 10^7$ years, much shorter than a Hubble time, the effect 
of an AGN on gas at large distances may be modeled as a blast 
wave. Adopting the Sedov-Taylor solution, we can write  
the postshock entropy as
\begin{equation}
S_{\rm s}=1.8\times 10^4 \kevcm2 E_{60} \delta_{\rm s}^{-5/3}
(1+z)^{-5}R_{\rm s,Mpc}^{-3}\,,
\end{equation}
where $E_{60}$ is the kinetic energy in an AGN outflow, E,  
in units of $10^{60}{\rm ergs}$, $\delta_{\rm s}$ is 
the overdensity of the surrounding medium defined as 
$\delta_{\rm s}=\rho_b/\bar{\rho}_b$, and 
$R_{\rm s,Mpc}$ is the shock radius in Mpc and is   
given by the Sedov solution as 
\begin{equation}
R_{\rm s}=1.7{\rm Mpc}\, E_{60}^{1/5}\delta_{\rm s}^{-1/5} 
(1+z)^{-3/5} t_{\rm Gyr}^{2/5},
\end{equation}
with $t_{\rm Gyr}$, the evolution time of the shock in 
gigayears. Following Scannapieco \& Oh (2004), we assume that $E$ is 
a constant fraction of the bolometric luminosity, $L_{\rm bol}$: 
$E\approx \epsilon_{\rm k} L_{\rm bol} t_{\rm AGN}$, where we take 
$\epsilon_{\rm k} \sim 0.05$. The AGN life time, $t_{\rm AGN}$, 
is assumed to be proportional to the dynamical time of a 
virialized halo, and is approximated by
$t_{\rm AGN}\approx 5.2\times 10^7 {\rm yr}\, (1+z)^{-3/2}$
(e.g. Barkana \& Loeb 2001). We assume the bolometric luminosity 
related to the luminosity in the blue band, 
$L_{\rm bol}=10.4L_{\rm B}$, as suggested by observations (Elvis \etal 1994).
Assuming that an AGN shines at its Eddington luminosity, 
one can relate the B-band luminosity to the mass of the black hole 
($M_{\rm BH}$) as 
$L_{\rm Edd,B}=5.73\times 10^3 L_{\odot}M_{\odot}^{-1} M_{\rm BH}$
(e.g. Willot, McLure \& Jarvis 2003). 
We use the observed relation between $M_{\rm BH}$ and the 
circular velocity $v_{\rm c}$ obtained by Ferrarese \etal (2002),
\begin{equation}
M_{\rm BH}=1.4\times10^8 M_{\odot}F
\left({v_{\rm c} \over 300{\rm km s^{-1}}} \right)^5,
\end{equation}
to estimate the black hole mass, where $F$ is a model 
parameter taken to be $0.6$, as in Wyithe \& Loeb (2003).
With all these assumptions, one can estimate $S_{\rm ph, 12}\sim 7 \kevcm2$ 
at $z=2$, assuming $\delta_{\rm s}=1$.

\begin{figure}
\centerline{\psfig{figure=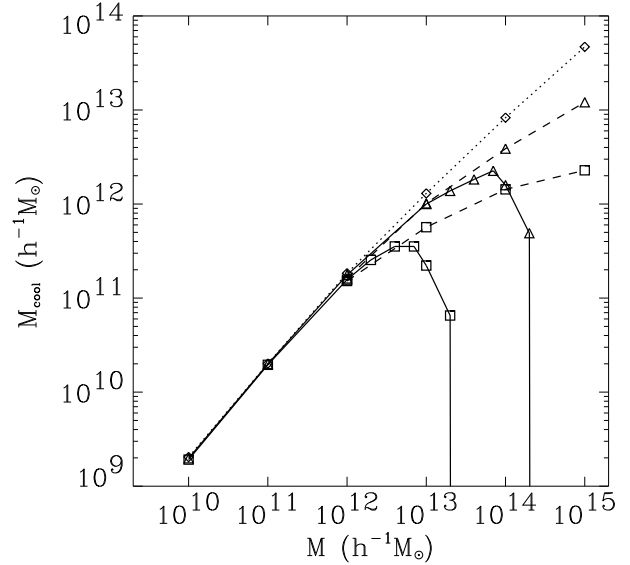,width=\hssize}}
\caption{Cooled gas mass versus halo mass in simulations 
assuming AGN preheating. The squares show results 
assuming $S_{\rm ph,12}=7 \kevcm2$, and the triangles show 
results assuming $S_{\rm ph,12}=3.5 \kevcm2$.  
Symbols connected by solid lines are results assuming 
a single-phase medium, and symbols connected by dashed 
curves are results assuming the ``cooling tail'' model of 
the multi-phase medium.  For comparison, diamonds connected 
by dotted lines are the results assuming no preheating.
}
\label{fig:mcool2}
\end{figure}

We first carry out single-phase simulations similar to those 
presented in Section \ref{sec:cooling}, but with the preheating entropy 
set by (\ref{AGN_sph}) with $S_{\rm ph,12}=7 \kevcm2$. For  
comparison, we also consider another sequence 
with $S_{\rm ph,12}=3.5 \kevcm2$. Figure\,\ref{fig:mcool2}
shows the mass of cooled gas at $z=0$ as a function of halo 
mass. It is clear that the suppression of gas cooling by AGN preheating
is more important in more massive halos. This result can be 
understood using Figure\,\ref{fig:sc}. Since the 
AGN preheating entropy considered here is proportional to the 
virial entropy of the halo, the effect of preheating is more important
in more massive halos. In this model, there is an upper-mass limit 
for the cooled gas mass, which is about $\sim 3\times 10^{11}\hMsun$
assuming $S_{\rm ph,12}=7 \kevcm2$, and about 
$\sim 2\times 10^{12}\hMsun$ assuming $S_{\rm ph,12}=3.5\kevcm2$. 
This maximum mass occurs at a halo mass of $M\sim 4\times 10^{12}$
for $S_{\rm ph,12}=7 \kevcm2$, and at $M\sim 5\times 10^{13}$
for $S_{\rm ph,12}=3.5 \kevcm2$. Thus, for a given preheating
amplitude $S_{\rm ph,12}$, significant gas cooling in halos above 
a certain mass can only occur in progenitors before the formation 
of the central AGN. As soon as the central AGN has formed, gas
cooling is effectively suppressed by preheating. 

This result has significant implications.
Observations of AGNs suggest that there is an upper limit for quasar 
luminosity (Fan \etal 2003), which corresponds to 
a black hole with a mass of $\sim 10^{9}\Msun$ assuming it radiates with an 
Eddington luminosity. This mass is approximately equal to 
that of the central black hole in a halo with a circular 
velocity $v_{\rm c}\sim 470 {\rm km\,s^{-1}}$ (corresponding to a halo mass of 
$10^{13.5}\Msun$ at the present time), using the observed 
$M_{\rm BH}-v_{\rm c}$ relation (Wyithe \& Loeb 2003). 
If the mass of a central black hole scales with its host halo as
the $M_{\rm BH}-v_{\rm c}$ relation for all dark matter halos, 
such an upper limit for black hole mass would not be expected.   
Thus, there must be a process that can prevent the growth of  
central black holes in massive halos. The AGN preheating 
model considered here may alleviate this difficulty. 
Before the formation of the supermassive black hole, gas 
can cool effectively and the mass of the central black hole 
grows rapidly. However, as soon as the mass of the central
black hole reaches a value of $\sim 10^{9}\Msun$, gas cooling is 
effectively suppressed and the black hole stops growing. The same 
process also prevents the growth of the central galaxy, so that 
the 1-d velocity dispersion of the central galaxy cannot be much 
larger than $330 {\rm km\,s^{-1}}$. This may explain why
not many galaxies can have velocity dispersions much larger than 
this (e.g. Sheth et al. 2003). 

  There are, however, a number of uncertainties in this 
conclusion. First of all, if the growth of black hole mass is 
indeed saturated at a fixed value, the preheating produced by the 
AGN would become  unimportant as the dark matter halo grows 
and the corresponding $\sv$ becomes larger than the entropy 
generated by the AGN.  Cooling will then resume in halos with 
$v_{\rm c}\gg 330{\rm km\,s^{-1}}$. To prevent 
further cooling in massive halos after the formation of 
SMBs, one may still need energy sources to balance the 
radiative cooling. One possibility here is the feedback 
from the central galaxy in ``radio'' mode. Although the energy 
feedback in such a mode is much weaker than that in the ``quasar'' 
mode, it may be sufficient to balance radiative cooling in 
massive halos (e.g. Croton et al. 2006). Another uncertainty 
is the possible development of a multi-phase medium in massive halos, which 
may promote additional radiative cooling. To demonstrate this,   
we run multi-phase simulations, assuming the ``cooling tail''
model, for halos more massive than $10^{12}\hMsun$. The dashed 
curves in Figure\,\ref{fig:mcool2} show the mass of cooled gas at $z=0$ 
as a function of halo mass for this model. Now, there is no 
sharp truncation in the cold gas mass, although the total 
cold gas mass is still reduced relative to the non-preheating model.
Unfortunately, it is unclear whether a significant multi-phase 
medium can develop in massive halos. Indeed, in order for the 
`radio-mode' feedback to work in massive halos, the formation 
of multi-phase media has to be suppressed. 

\subsection{Low-mass halos in a warm intergalactic medium}

As one can see from Figure\,\ref{fig:mcool2}, AGN preheating has little 
effect on small halos, because the black holes in these systems 
are small and the cooling is efficient. Thus, AGN preheating 
{\it cannot} suppress star formation in low-mass halos 
to explain the flat faint-end slope of the galaxy luminosity 
function. Some other processes must be at work in small halos.
In this subsection, we consider a model in which all 
small halos ($M<10^{12}\Msun$) are assumed to be embedded in 
a warm medium with an initial specific entropy $\sph \sim 10\kevcm2$.
This model is motivated by the proposal of Mo \etal (2005)
that low-mass halos at low redshift may be surrounded 
by a warm intergalactic medium produced by gravitational 
pancaking. Another possibility for such preheating is supernova 
explosions associated with the formation of halo stars,  
bulge stars, and stars in the thick disk. 

\begin{figure}
\centerline{\psfig{figure=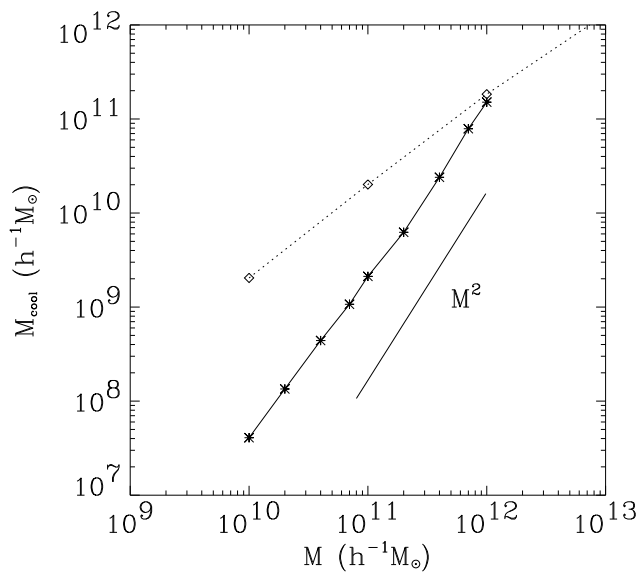,width=\hssize}}
\caption {
The mass of cooled gas versus halo mass for a model 
in which the intergalactic gas is preheated to a specific 
entropy $\sph=8.0\kevcm2$ (stars).
In this case, the cooled gas mass scales with halo mass 
roughly as $M^2$. For comparison, the diamonds show the results
for a model without preheating. 
}
\label{fig:mcool3}
\end{figure}

  We consider a case where $\sph=8\kevcm2$.
Since such a low value of $\sph$ has little effect on 
halos with $M>10^{12}\Msun$, we concentrate on smaller halos. 
Figure\,\ref{fig:mcool3} shows the mass
of cooled gas, $M_{\rm cool}$, as a function of halo mass. 
For low mass halos, $M_{\rm cool}$ scales with halo mass roughly 
as $M^2$. This can be understood as follows. As we have seen in 
Section \ref{sec:adiabatic}, if $\sph\geq \sv$, as is the case 
for low-mass halos, the fraction of 
accreted gas scales as $f_{\rm acc}=(\sph /\sv)^{-3/2}$. 
Since at a fixed redshift $\sv\propto T_{\rm v}$ and 
$T_{\rm v}\propto M^{2/3}$, the mass of the accreted gas 
$M_{\rm acc}=f_{\rm acc}f_bM \propto M^{2}$ for a constant 
preheating entropy. In small halos almost all the accreted 
gas can cool, so we have $M_{\rm cool}\propto M^2$. This result 
supports the assumption made by Mo \etal (2005) about the 
$M_{\rm cool}$ - $M$ relation for low-mass halos accreting gas 
in a medium preheated by the formation of pancakes. 
As shown in Mo \etal (2005), such a $M_{\rm cool}$ - $M$ 
relation is required to explain bothe the faint-end
slope of the galaxy luminosity function and the HI-mass function 
of galaxies. It should be noted that the multi-phase model 
does not change the cold gas mass in these halos. 
As we have demonstrated in Section~\ref{sec:cooling}, the cooling 
is always efficient in small halos. Hence, once the gas is accreted 
into low-mass halos, it can cool no matter it was in high-density 
or low-density phases. Note, however, that the assembly history 
of the cold gas into the central disk may  depend significantly on 
whether the medium is single-phase or multi-phase, as discussed
in Section \ref{sec:multi}.
 
\section{Summary}
\label{sec:sum}

 In this paper, we examine the effects of preheating 
on gas cooling and accretion into dark matter halos, 
using a 1-D hydrodynamical code and realistic halo formation 
histories. We explore the behavior of the preheated gas 
by varying the virial mass of the 
halo and the amount of preheating. We find that preheating 
can both prevent gas accretion into dark matter halos and reduce the 
radiative cooling of that gas. If the preheating entropy 
is larger than the virial entropy of the halo, the fraction of the 
gas that can be accreted into a halo scales roughly as 
$f_g\propto (\sph/\sv)^{-3/2}$. For low-mass halos, almost all the 
accreted gas can cool, and so preheating reduces the total 
amount of cold gas that can be assembled into the central galaxy 
only by reducing the total amount of gas that can be accreted by 
the halo. For massive halos, on the other hand, 
preheating not only reduces the amount of gas that can 
be accreted, but also suppresses radiative cooling. 
We also examine the effects of a multi-phase medium on the 
cooling of halo gas and find that the cooling history of 
multi-phase gas is very different from that of 
a single-phase medium model. In a multi-phase medium, gas cooling   
can occur in the dense phases soon after the gas is accreted 
into the halo, in contrast to a single-phase medium 
where gas cooling is delayed. Also, 
in a single-phase medium, gas cooling in general proceeds in 
an inside-out fashion, while in a multi-phase medium gas can 
cool, at a given time, over a wide range of radii.

We consider two examples where preheating may have played important 
roles in galaxy formation and evolution. First, we examine a case 
where the gas associated with a dark matter halo is preheated by an 
AGN outflow powered by a central supermassive black
hole. Using assumptions made in recent quasar feedback models, 
we demonstrate that such preheating can effectively suppress gas 
cooling in halos with masses above $\sim 10^{13}h^{-1}{\rm M}_\odot$,  
leading to an upper mass limit of about $10^{12}\hMsun$ for the central 
galaxy that can form in such a massive halo. This may be the reason 
why the galaxy luminosity (mass) function has a sharp break at the 
bright (massive) end. This suppression of gas cooling in massive 
halos may also be responsible for the observed bimodality in the 
colour-magnitude diagram. Observations show that the blue sequence,
which consists of galaxies with significant star formation,  
is truncated at the bright end, with very few blue galaxies 
having stellar masses exceeding $10^{11}{\rm M}_\odot$. The
more massive galaxies are all red, with little current star 
formation. This suggests that there must be some process 
to quench star formation in these massive galaxies. 
Recent semi-analytical models of galaxy formation including AGN feedback 
show that the bimodality in the colour distribution and the truncation of 
the blue sequence can be reproduced if the feedback from the central 
galaxy is in the ``radio mode'' (e.g. Croton et al 2005; Cattaneo et al 2006). 
However, in order for the ``radio-mode'' feedback to be effective, 
the halo gas must be in a diffuse hot form. In the scenario we 
consider here, the halo gas is preheated by the feedback from the 
central AGN in the ``quasar mode'', and any subsequent cooling is 
suppressed in the preheated medium. This may help create the hot 
diffuse halos within which the ``radio mode'' feedback can act 
effectively. 

 The second example we consider is the case where the intergalactic
medium is assumed to be preheated to a constant entropy level. This model 
is motivated by the scenario outlined in Mo \etal (2005) where such 
preheating may help explain the observed stellar- and HI-mass functions
of galaxies at the low-mass end in current CDM models. Our simulations confirm that 
such preheating can indeed produce the cold-gas-mass/halo-mass 
relation, $M_{\rm cool} \propto M^2$, as is required to 
match the observed stellar- and HI-mass functions in the current
CDM theory.

  The results obtained in this paper demonstrate that preheating may 
have played an important role in galaxy formation. To explore 
the observational consequences of the preheating scenario, one 
needs to combine the results obtained here with the dark matter 
halo population predicted by current CDM models of structure 
formation, to quantify how preheating affects the properties of 
the galaxy population and of the intergalactic medium. The best 
approach is perhaps through semi-analytic modeling, and we 
plan to do this in a future paper.   

\section*{Acknowledgment}
We thank David Weinberg and Anne Thoul for providing their 1-d hydrodynamical
code that helped the development of the code used in this paper.
We thank Neal Katz for reading the manuscript and many beneficial comments. 
We thank Avi Loeb for useful discussion.
HJM would like to acknowledge the support of NSF AST-0607535, 
NASA AISR-126270 and NSF IIS-0611948. 


\label{lastpage}

\end{document}

%% file: macro1.tex
\newcommand{\etal}{{et al.~}}

\newcommand{\Msun}{\>{\rm M_{\odot}}}

\newcommand{\beq}{\begin{equation}}
\newcommand{\eeq}{\end{equation}}

\newcommand{\rmd}{{\rm d}}

\def\sph{S_{\rm ph}}
\def\sv{S_{\rm v}}

\def\Msun{\,\rm M_{\odot}}
\def\hMsun{\,h^{-1}\Msun}
\def\kevcm2{\,{\rm keV\,cm}^2}

\newdimen\hssize
\newdimen\hdsize 